\documentclass[onecolumn,a4paper,fleqn,usenatbib]{mnras}

\usepackage{newtxtext,newtxmath,mymath}

\usepackage[T1]{fontenc}
\usepackage{ae,aecompl}

\usepackage{graphicx}	
\usepackage{amsmath}	

\newtheorem{theorem}{Theorem}

\newcommand{\erf}{\mathop{\mathrm{erf}}}

\title[Asymptotics of the Cosmic Power Spectrum]{On the asymptotic behaviour of cosmic density-fluctuation power spectra}

\author[S. Konrad and M. Bartelmann]{
Sara Konrad$^{1}$\thanks{fz002@uni-heidelberg.de}
and Matthias Bartelmann$^1$
\\
$^1$ Institute for Theoretical Physics, Heidelberg University, Germany\\
}

\date{Accepted XXX. Received YYY; in original form ZZZ}

\pubyear{2022}

\begin{document}
\label{firstpage}
\pagerange{\pageref{firstpage}--\pageref{lastpage}}
\maketitle

\begin{abstract}
We study the small-scale asymptotic behaviour of the cosmic density-fluctuation power spectrum in the Zel'dovich approximation. For doing so, we extend Laplace's method in arbitrary dimensions and use it to prove that this power spectrum necessarily develops an asymptotic tail proportional to $k^{-3}$, irrespective of the cosmological model and the power spectrum of the initial matter distribution. The exponent $-3$ is set only by the number of spatial dimensions. We derive the complete asymptotic series of the power spectrum and compare the leading- and next-to-leading-order terms to derive characteristic scales for the onset of non-linear structure formation, independent of the cosmological model and the type of dark matter. Combined with earlier results on the mean-field approximation for including particle interactions, this asymptotic behaviour is likely to remain valid beyond the Zel'dovich approximation. Due to their insensitivity to cosmological assumptions, our results are generally applicable to particle distributions with positions and momenta drawn from a Gaussian random field. We discuss an analytically solvable toy model to further illustrate the formation of the $k^{-3}$ asymptotic tail.
\end{abstract}




\section{Introduction}
\label{sec:1}
Numerical simulations reveal that cosmic structures develop universal properties, for which the radial density profiles of gravitationally-bound objects are an important example \cite[for recent work]{Springel2005, 2017MNRAS.469.1824X, Springel2018, 2019MNRAS.484..476C, Vogelsberger2020}. Where does this universality originate from? This question can be addressed in the framework of an analytic theory of cosmic structure formation. We use previous results obtained with the kinetic field theory of cosmic structure formation (KFT) \cite{Bartelmann2016, Bartelmann2017, Bartelmann2019} to study the asymptotic behaviour of the power spectrum of cosmic density fluctuations in the limit of small scales, or larger wave numbers, $k\to\infty$.

The central object of KFT is the generating functional for bundles of classical particle trajectories on the expanding cosmological background space-time. This generating functional incorporates a probability distribution for the intial particle distribution in phase space, a Green's function or propagator $g_{qp}(t,t')$, solving the Hamiltonian equations of motion, and an interaction term relative to the propagator chosen \cite{Bartelmann2015, 2021ScPP...10..153B}.

Initial particle positions and momenta can be drawn from a velocity potential $\psi$ because the initial velocity field can be considered curl-free, and density fluctuations are related to the fluctuations of this potential by the continuity equation. Assuming that $\psi$ is a Gaussian random field, the initial particle distribution can be statistically completely characterized by the power spectrum $P_\psi^\mathrm{(i)}$ of the velocity potential \cite{Bartelmann2016}.

One specific and advantageous choice for the particle propagator is given by the Zel'dovich approximation \cite{1970A&A.....5...84Z}. It asserts that particles move approximately along straight lines in a time coordinate set by the linear growth factor $D_+$ of cosmic density fluctuations. Large-scale gravitational interaction is contained in the Zel'dovich approximation because initial particle momenta are correlated in such a way that particle streams are convergent at matter overdensities, and divergent at underdensities. We have shown in earlier work how particle interactions relative to the Zel'dovich trajectories can be described in a mean-field approximation \cite{2021ScPP...10..153B}.

In this paper, we study the asymptotic evolution of the free KFT power spectrum on small scales, i.e.\ large wave numbers $k$. By this, we mean the power spectrum of the cosmic density field, decomposed into classical particles underlying Hamiltonian dynamics, neglecting particle-particle interactions and modeling particle trajectories with the Zel'dovich approximation. The Zel'dovich power spectrum is non-linear in the sense that arbiratrily large density fluctuations can occur in the Zel'dovich approximation. Particle trajectories, which are linear in the time coordinate in the Zel'dovich approximation, enter non-linearly into the power spectrum. Thus, the Zel'dovich approximation captures part of the non-linear structure evolution. On the basis of our results, it should be possible to analyze the origin of the observed universality of dark matter halo density profiles, which is however not subject of this but of future work.

We have shown in earlier papers \cite{Bartelmann2016, Bartelmann2017} that the free KFT power spectrum is given by
\begin{equation}
  \mathcal{P}(k,t) = \E^{-Q}\int\D^3q\left(
    \E^{g_{qp}^2(t,0)\,\vec k^\top\hat{C}_{pp}(\vec q\,)\vec k}-1
  \right)\E^{\I\vec k\cdot\vec q}
\label{eq:1}
\end{equation}
with the initial auto-correlation matrix
\begin{align}
  \hat{C}_{pp}\left(\vec q\,\right) = \int_{k'}\left(
    \vec k'\otimes\vec k'\,
  \right)P_\psi^\mathrm{(i)}(k')\,\E^{\I \vec k'\cdot\vec q}
\label{eq:2}
\end{align}
of particle momenta. The exponent $Q$ is
\begin{equation}
  Q = \frac{\sigma_1^2}{3}\,k^2\,g_{qp}^2(t,0)\;,
\label{eq:3}
\end{equation}
where $\sigma_1^2$ is one of the moments
\begin{equation}
  \sigma_n^2 = \frac{1}{2\pi^2}\int_0^\infty\D k\,k^{2n+2}\,
  P_\psi^\mathrm{(i)}(k)
\label{eq:4}
\end{equation}
of the power spectrum $P_\psi^\mathrm{(i)}$ of the initial velocity potential $\psi$. It is important for the present derivation of our results that $\sigma_2$ is finite. If the initial power spectrum does not fall off steeply enough for large $k$, we introduce an exponential cut-off at a large wave number $k_\mathrm{s}$ to be specified below. Via Poisson's equation, this spectrum is related to the initial density-fluctuation power spectrum $P_\delta^\mathrm{(i)}$ by
\begin{equation}
  P_\psi^\mathrm{(i)}(k) = k^{-4}\,P_\delta^\mathrm{(i)}(k)\;.
\label{eq:5}
\end{equation} 

Equation (\ref{eq:1}) has been derived by assuming (i) straight particle trajectories, (ii) a curl-free initial velocity field with velocity potential $\psi$, (iii) an initial Gaussian random field, and (iv) taking only the initial momentum-momentum correlations into account, i.e.\ neglecting density-density and density-momentum correlations. The free power spectrum in KFT is equivalent to the non-linear Zel'dovich power spectrum when we specify $g_{qp}$ to be the Zel'dovich propagator $g_{qp}^\mathrm{(Z)}(t,0) = D_+(t)-D_+(t^{\text{(i)}})$, where $D_+$ is the linear growth factor. At the same time, it is the unperturbed zeroth-order power spectrum in KFT and thus the starting point for calculating the non-linear power spectrum from KFT perturbation theory or mean-field approaches \cite{Bartelmann2016, 2021ScPP...10..153B}.

The remainder of this paper has three parts. In Sect.~\ref{sec:2}, we present our result for the leading-order, asymptotic behaviour of the free KFT power spectrum $\mathcal{P}$. Our main result there will be that $\mathcal{P}\sim k^{-3}$ for $k\to\infty$, where the exponent $-3$ depends neither on the cosmological model nor on the form of the initial power spectrum. We discuss cosmological consequences in Sect.~\ref{sec:3}, deriving a time scale for the onset of non-linear structure growth on small scales and a maximum amplitude for the free power spectrum which are both universal in the sense that they depend neither on the cosmological model nor on the power spectrum of the initial state, and thus on the dark-matter model.

In Sect.~\ref{sec:4}, we derive the entire asymptotic series of the free power spectrum. We show there that $m$-th order terms in this series, $m\in\mathbb{N}_0$, fall off asymptotically like $k^{-3-2m}$, equally independent of cosmology and the type of dark matter. Also in Sect.~\ref{sec:4}, we introduce an analytically solvable toy model which may be useful to clarify the origin of the asymptotic $k^{-3}$ tail developed by the free power spectrum. In Sect.~\ref{sec:5}, we summarize our paper and discuss our conclusions. Detailed calculations are collected in the Appendix for the main text to be more readable.

\section{Small-scale asymptotics of the free power spectrum}
\label{sec:2}

We derive here the leading-order asymptotics of the free power spectrum for large wave numbers, $k\to\infty$. Since $k$ appears in the exponent of the integrand in (\ref{eq:1}) with a different power than in the oscillating phase factor, no standard asymptotic method exists that could be applied directly. We thus develop for this case a modification of Laplace's method for $d$-dimensional integrals. We motivate our method in the main text and give a detailed derivation in Appendix \ref{sec:A}, involving Morse's lemma and a resummation of Laplace's method.

For evaluating the real exponent in (\ref{eq:1}), we rotate the coordinate frame such that the $z$ axis points into $\vec k$ direction and write
\begin{equation}
  g_{qp}^2(t,0)\vec k^\top\hat{C}_{pp}(\vec q\,)\vec k =
  -g_{qp}^2(t,0)k^2f\left(\vec q\,\right)\;,
\label{eq:6}
\end{equation}
defining the function
\begin{equation}
  f\left(\vec q\,\right) = -\int_{k'}k_z'^2P_\psi(k')\,
  \E^{\I\vec k'\cdot\vec q}\;.
\label{eq:7}
\end{equation}
The free power spectrum then reads
\begin{equation}
  \mathcal{P}(k,t) = \E^{-Q}\left(
    \int\D^3q\,\E^{-g_{qp}^2 k^2 f(\vec q\,)}\,\E^{\I\vec k\cdot\vec q}-
    (2\pi)^3\delta_\mathrm{D}(\vec k\,)
  \right)\;.
\label{eq:8}
\end{equation}
We neglect the delta distribution in the following because we focus on $k>0$.

The idea of our modification of Laplace's method is now the following: For $k\to\infty$, the integrand sharply peaks where $f(\vec q\,)$ has an isolated global minimum, and it is exponentially suppressed away from it. For the result of the integral, the most relevant part of the integrand is then the local environment of the minimum of $f(\vec q\,)$. To derive the asymptotics of $\mathcal{P}(k,t)$ for $k\to\infty$, we thus expand $f(\vec q\,)$ up to quadratic order in $\vec q$ and then analytically integrate this new integral.

Since the integrand in (\ref{eq:7}) except for the Fourier phase is strictly positive, the function $f(\vec q\,)$ has an isolated global minimum at $\vec q = 0$,
\begin{align}
  \left|f\left(\vec q\,\right)\right| &\le \int_{k'}k_z'^2P_\psi(k') =
  \frac{1}{(2\pi)^2}\int_0^\infty\D k'k'^4P_\psi(k')\int_{-1}^1\D\mu'\mu'^2
  \nonumber\\ &=
  \frac{\sigma_1^2}{3} = -f(0)\;,
\label{eq:9}
\end{align} 
with a vanishing gradient there, $\vec\nabla f(\vec q\,)|_{\vec q=0}=0$. The Hessian $A$ of $f(\vec q\,)$ at $\vec q = 0$ has the components
\begin{equation}
  A_{ij} = \left.
    \frac{\partial^2f\left(\vec q\,\right)}{\partial q_i\partial q_j}
  \right|_{\vec q=0} = 
  \int_{k'}k_z'^2\,k'_ik'_j\,P_\psi(k') =
  \frac{\sigma_2^2}{15}\left(
    \delta_{ij}+2\delta_{iz}\delta_{jz}
  \right)
\label{eq:10}
\end{equation}
in terms of the moment $\sigma_2^2$. Since $A$ is diagonal and has only positive entries, it is positive definite and has the determinant
\begin{equation}
  \det A = 3\left(\frac{\sigma_2^2}{15}\right)^3 > 0\;,
\label{eq:11}
\end{equation}
and its inverse has the components
\begin{equation}
  A^{-1}_{ij} = \frac{15}{\sigma_2^2}\left(
    \delta_{ij}-\frac{2}{3}\delta_{iz}\delta_{jz}
  \right)\;.
\label{eq:12}
\end{equation} 

These results show that all requirements listed in Appendix \ref{sec:A}.3 are met, except for one. It is left to show that $f(\vec q\,)$ is quadratically integrable in $\mathbb{R}^3$. This can be seen as follows: $f(\vec q\,)$ is the inverse Fourier transform of the function $F(\vec k)= k_z^2 P_{\psi}(k)$. This function is quadratically integrable in $\mathbb{R}^3$ because it falls of fast enough for large $k$, while $k^4 P^2_{\psi}(k) \propto k^{2n_\mathrm{s}-4}$ with $n_\mathrm{s}\lesssim1$ for small $k$. Since the Fourier transform maps quadratically integrable functions into themselves, $f(\vec q\,)$ is also quadratically integrable.

We can thus apply our result (\ref{eq:60}) of Appendix \ref{sec:A} with $s = 2$ and $d = 3$. Using the expression for $f(0)$ given in (\ref{eq:9}), the determinant (\ref{eq:11}) of $A$, the $zz$-component of its inverse from (\ref{eq:12}), denoting the linearly evolved $n$-th moment of the initial velocity potential field by
\begin{equation}
  \tau^2_n(t)\coloneqq g_{qp}^2(t,0)\sigma_n^2
\label{eq:13}
\end{equation}
and introducing the abbreviation
\begin{equation}
  \Sigma(t) = \frac{5}{2\tau_2^2(t)}\;,
\label{eq:14}
\end{equation} 
we can state that the leading-order asymptotics of the free power spectrum for $k\to\infty$ is
\begin{equation}
  \mathcal{P}(k,t) \sim \frac{3\,(4\pi)^{3/2} }{k^3}\Sigma^{3/2}(t)\,
  \E^{-\Sigma(t)}
\label{eq:15}
\end{equation}
for $k\to\infty$. This result proves that the free power spectrum inevitably falls off asymptotically like $k^{-3}$. This result is remarkably independent of the initial power spectrum. The only condition is that the moment $\sigma_2^2$ exists. The apparent exponential damping by $\E^{-Q}$ in (\ref{eq:1}) is completely canceled by the oscillating integral.

\section{Implications for cosmic structure formation}
\label{sec:3}

In the preceding Section, we proved that the power spectrum for particles on free trajectories with initially correlated momenta drawn from a Gaussian random velocity potential always develops a $k^{-3}$ tail for large wave numbers $k$. This universal power law occurs independently of the cosmological model and the form of the initial power spectrum at large $k$ values. The exponent $-3$ is solely determined by the number of spatial dimensions and has nothing to do with the dark-matter species assumed.

\subsection{Comparison of analytic to numerical results}

\begin{figure}
  \includegraphics[width=\hsize]{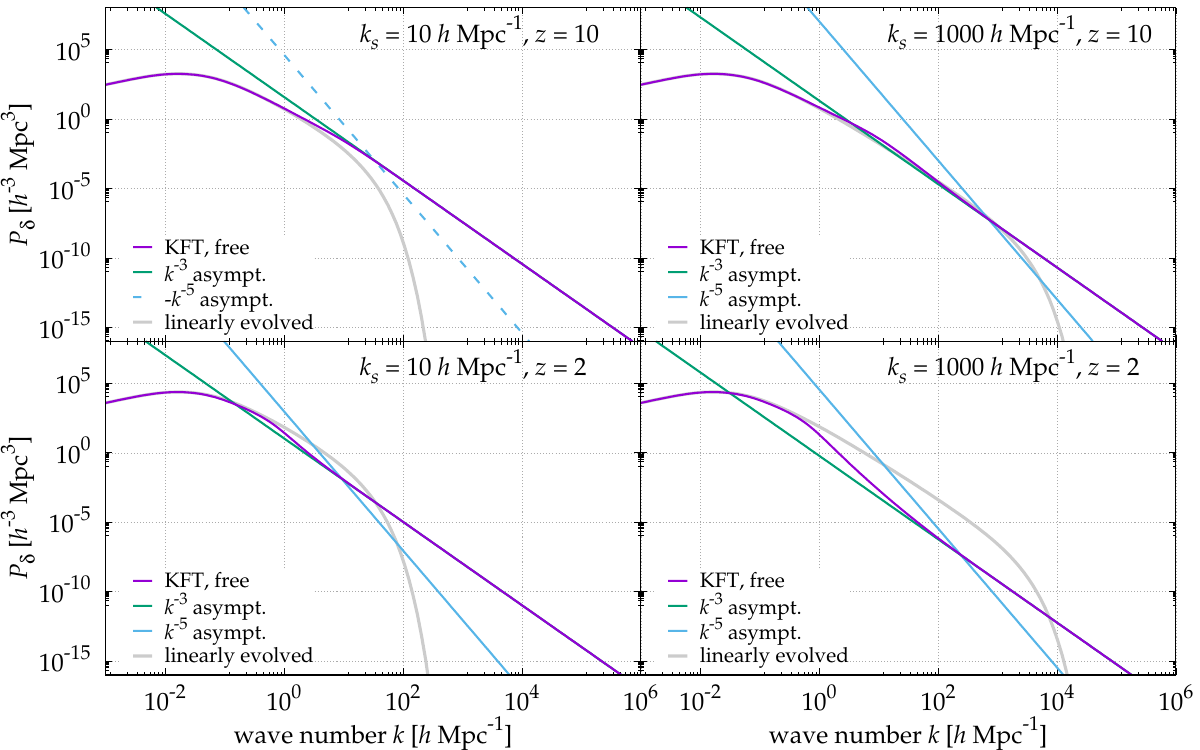}
\caption{At large wave numbers, the $k^{-3}$ asymptotics (\ref{eq:15}) (green lines) is in perfect agreement with the numerical evaluation of the free power spectrum (\ref{eq:1}) (purple lines) for early (top) and late (bottom) times as well as for large (left) and small (right) smoothing scales. The second order of the asymptotic expansion (\ref{eq:28}) (blue lines) predicts the scale of the onset of the $k^{-3}$ behavior at those wave vectors where the $k^{-5}$ term falls below the $k^{-3}$ term. The blue dashed line in the top-left panel indicates that the $k^{-5}$ term is negative. The linearly evolved smoothed power spectra (\ref{eq:16}, grey lines) are drawn for comparison. The cosmological parameters used are listed in Tab.~\ref{tab:1}.}
\label{fig:1}
\end{figure}

To demonstrate the validity of our asymptotic expansion, we proceed to compare our analytic with numerical results. For the initial density-fluctuation power spectrum $P^\mathrm{(i)}_\delta$, we choose the cold dark matter power spectrum from \cite{Bardeen1986} and multiply it with an exponential regulator with smoothing scale $k_\mathrm{s}$. Figure~\ref{fig:1} shows the free power spectrum (purple lines), together with the leading-order $k^{-3}$ term (\ref{eq:15}, green lines) and the $k^{-5}$ term appearing at the next order in the full asymptotic expansion derived below in (\ref{eq:28}, blue lines), for two different redshifts ($z = 2$ and $10$) and two different smoothing scales ($k_\mathrm{s}=10$ and $1000\,h\,\mathrm{Mpc}^{-1}$). We shall discuss free power spectra at redshift $z = 0$ later. For comparison, we also show the linearly evolved smoothed power spectra (grey lines),
\begin{equation}
  P^\mathrm{(lin)}_\delta(k) = g_{qp}^2(t,0)\,P^{\text{(i)}}_\delta(k)\;.
\label{eq:16}
\end{equation}
To relate the propagator to redshift, we choose the Zel'dovich propagator $g_{qp}^\mathrm{(Z)}(t,0) = D_+(t)-D_+(t^{\text{(i)}})$ and set the initial time $t^\mathrm{(i)} = 0$. For calculating the linear growth function $D_+$, we use the cosmological parameters listed in Tab.~\ref{tab:1}.

\begin{table}
\caption{Cosmological parameters used where needed}
\label{tab:1}
\begin{center}
\begin{tabular}{|c|c|c|c|c|c|}
  \hline
  $\Omega_{\Lambda 0}$ & $\Omega_{\mathrm{m}0}$ & $\Omega_{\mathrm{b}0}$ &
  $h$ & $\sigma_8$ & $n_s$ \\
  \hline
  $0.7$ & $0.3$ & $0.04$ & $0.7$ & $0.8$ & $0.96$ \\
  \hline
\end{tabular}
\end{center}
\end{table}

As expected, the free power spectrum converges in all four examples to the $k^{-3}$ asymptotics. A comparison with the linearly evolved power spectrum cut off at different wave numbers shows that the power on small scales is enormously increased: all formerly exponentially damped power spectra acquire a $k^{-3}$ tail. Moreover, the asymptotic behavior is independent of the first moment of the initial velocity potential $\sigma_1^2$, i.e.\ the initial velocity dispersion.

\subsection{Time-dependence of the amplitude}

The time-dependent amplitude of the leading-order asymptotics in (\ref{eq:15}), repeated as $\mathcal{P}^{(0)}(t)$ in (\ref{eq:28}), is a function of the product $g_{qp}^2(t,0)\sigma_2^2 = \tau_2^2$. Thus, for straight particle trajectories, $\tau_2$ acts as an effective time coordinate for structure growth on small scales. For small $\sigma_2$, small-scale structures evolve more slowly, while large values of $\sigma_2$ imply faster small-scale structure growth.

\begin{figure}
  \includegraphics[width=\hsize]{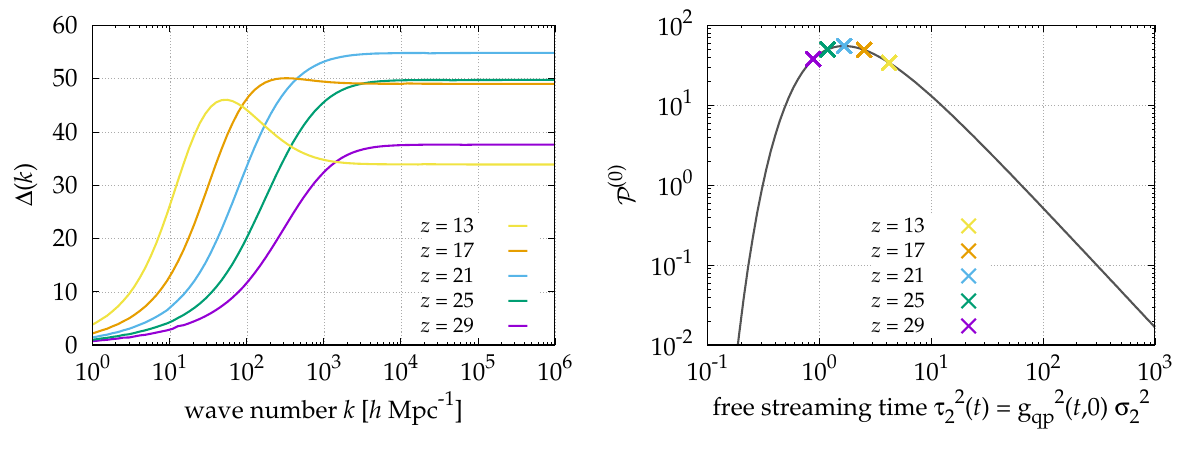}
\caption{Left: The free dimension-less power spectrum (\ref{eq:20}) for an initial CDM power spectrum exponentially cut off at $k_\mathrm{s} = 1000\,h\,\mathrm{Mpc}^{-1}$ is shown for different redshifts $z$. The amplitude rises from $z=29$ and reaches the universal maximum (\ref{eq:19}) at $z=21$. After the onset of re-expansion, the amplitude decreases while the $k^{-3}$ behavior persists. Right: The amplitude $\mathcal{P}^{(0)}$ of the leading-order asymptotic term of $\mathcal{P}$ as a function of the squared free-streaming time $\tau_2$ rises until it reaches the maximum value (\ref{eq:19}) at $\tau_2 = \sqrt{5/3}$, when stream-crossing occurs. After that, re-expansion leads to a decay of structures, and the amplitude decreases with $\tau_2^{-3}$. Colored points correspond to the asymptotic amplitudes shown in the left panel. The cosmological parameters used are listed in Tab.~\ref{tab:1}.}
\label{fig:2}
\end{figure}

The right-hand panel of Fig.~\ref{fig:2} shows the asymptotic amplitude $\mathcal{P}^{(0)}$ as a function of $\tau_2^2$, (grey line). Colored crosses mark the asymptotic amplitude of the power spectra shown in the left panel. The grey line shows that $\mathcal{P}^{(0)}$ first rises steeply during the free infall of structures until it reaches a maximum located at
\begin{equation}
  \tau_{2,\mathrm{max}} = \sqrt{\frac{5}{3}}\;.
\label{eq:17}
\end{equation}
This maximum indicates stream crossing at small scales because for larger values of $\tau_2$, re-expansion of structures sets in, such that the asymptotic amplitude $\mathcal{P}^{(0)}$ starts to decrease and eventually falls off like $\tau_2^{-3}$ for late times. Therefore, 
\begin{equation}
  g_{qp,\mathrm{max}} = \sqrt{\frac{5}{3\sigma_2^2}}\;
\label{eq:18}
\end{equation} 
implicitly sets the time of small-scale stream crossing. It is given in units of the linear growth factor, if the Zel'dovich propagator is chosen for $g_{qp}$. The amplitude of the leading-order asymptotic term reaches the maximum value
\begin{equation}
  \mathcal{P}_\mathrm{max}^{(0)} = 3\left(\frac{6\pi}{\e}\right)^{3/2}
  \approx 54.78\;,
\label{eq:19}
\end{equation}
which is universal for all cosmological models and all initial power spectra. The small-scale regime of the power spectrum will never grow beyond this amplitude when straight particle trajectories and initially correlated momenta are considered. Just the time for this universal maximum amplitude to be reached, given in (\ref{eq:18}) in units of of the linear growth factor, depends on $\sigma_2^2$.

In the left panel of Fig.~\ref{fig:2}, we show the free dimension-less power spectrum
\begin{equation}
  \Delta(k) = k^3\,\mathcal{P}(k)
\label{eq:20}
\end{equation} 
as a function of $\tau_2^2(t)$ for $k_\mathrm{s} = 1000\,h\,\mathrm{Mpc}^{-1}$ at five different redshifts. The dimension-less power spectra shown become constant for large values of $k$ because $\Delta(k)\to\mathcal{P}^{(0)}$ (\ref{eq:28}). Note that the plateaus show that the $k^{-3}$ asymptotics emerges already before stream crossing ($z=29$ and $25$ in Fig.~\ref{fig:2}) and persists after the onset of re-expansion ($z=17$ and $13$ in Fig.~\ref{fig:2}).

\subsection{Non-linear Zel'dovich power spectrum at redshift $z=0$}

On large scales, the non-linear Zel'dovich power spectrum is known to correspond to the linear power spectrum, which accurately describes the structure on the largest scales even today. However, it is also well known that the Zel'dovich approximation leads to a re-expansion of structures after stream crossing that goes along with decreasing power at small scales, as illustrated by Fig.~\ref{fig:2}.

In order to prevent this re-expansion, techniques like e.g.\ the adhesion approximation were applied; \cite[for example]{Weinberg1990}. Another method to lift the tail of the Zel'dovich power spectrum is to use a so-called truncated Zel'dovich power spectrum, e.g.\ \cite{Coles1993}. This means that the initial power spectrum is truncated in a way similar to the exponential cut-off. Thus the initial power is suppressed at wave numbers exceeding $k_\mathrm{s}$.

\begin{figure}
  \includegraphics[width=\hsize]{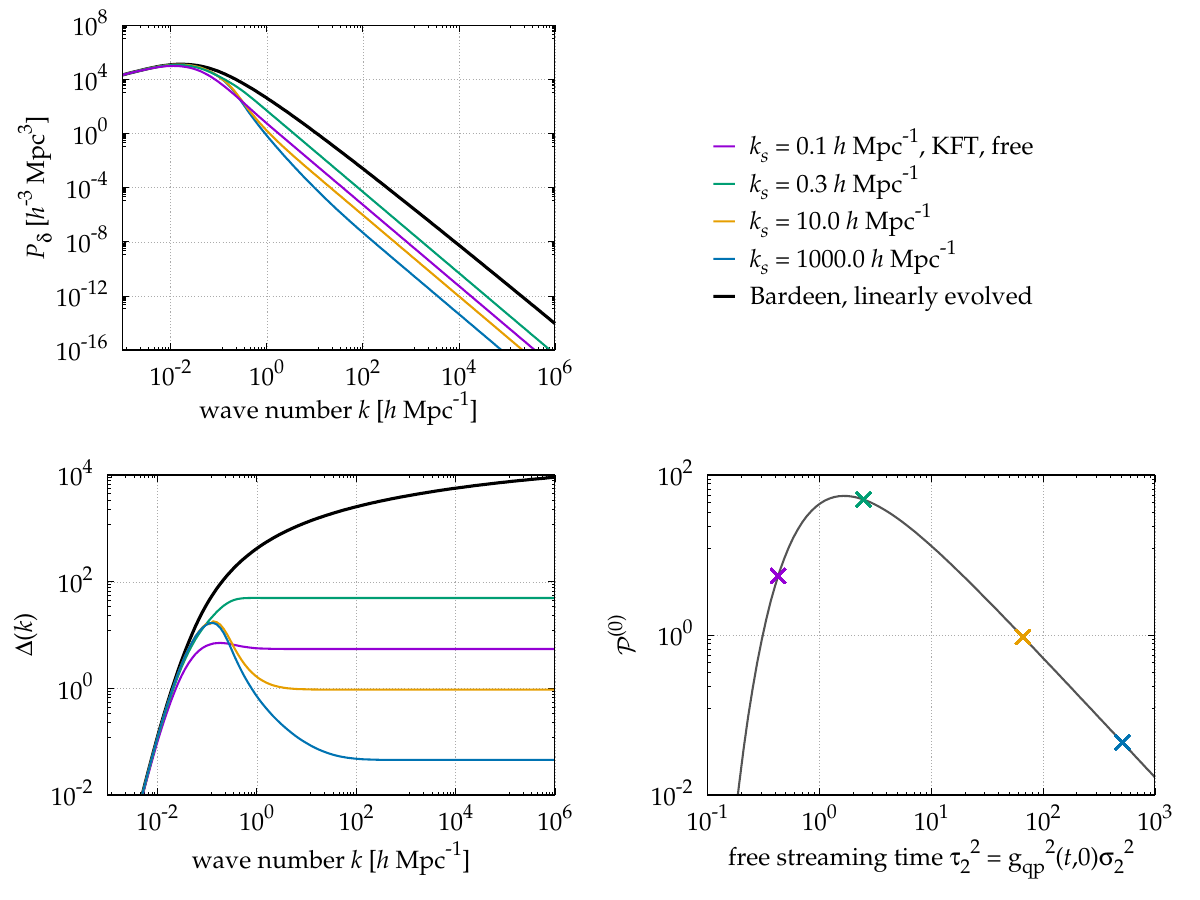}
\caption{The asymptotic amplitude evolves more slowly for smaller smoothing wave numbers. Top: The free power spectra (\ref{eq:1}) (colored lines) for different values of $k_s$ and the linearly evolved Bardeen power spectrum (black line) are shown at today's redshift $z=0$. The amplitudes at small scales differ by more than three orders of magnitude. Bottom left: The dimension-less power spectra as in the upper panel are shown. All free power spectra reach different plateau values for $k\gtrsim 100$ $h$ Mpc$^{-1}$. Bottom right: The asymptotic amplitude $\mathcal{P}^{(0)}$ (grey line) as a function of the free streaming time $\tau_2^2$ is shown as in the right panel of Fig.~\ref{fig:2}. Colored crosses indicate the plateau values of the dimension-less power spectra in the bottom left panel. For smaller $k_s$, the evolution of $\mathcal{P}^{(0)}$ proceeds slower, implying a slower growth of small-scale structures.  The cosmological parameters used are listed in Tab.~\ref{tab:1}.}
\label{fig:3}
\end{figure}

In Fig.~\ref{fig:3}, we show how this truncation, i.e.\ the suppression of small-scale fluctuations at different scales $k_\mathrm{s}$ affects the tail of the Zel'dovich power spectrum at $z=0$. In the top-left panel of Fig.~\ref{fig:3}, the free power spectra (colored lines) for four different cut-off scales are shown, together with the linearly evolved CDM power spectrum (black line). The tail of all four free power spectra stays below the linearly evolved CDM spectrum. In the bottom-left panel of Fig.~\ref{fig:3}, we show the same dimension-less power spectra that all reach plateau values for $k\gtrsim 100\,h\,\mathrm{Mpc}^{-1}$. In the bottom-right panel of Fig.~\ref{fig:3}, we show the amplitude $\mathcal{P}^{(0)}$ as a function of $\tau_2^2$ (grey line), where the colored crosses indicate the values for four example power spectra at $z=0$. The truncation at smaller wave numbers leads to smaller values of $\sigma_2^2$ causing the amplitude at scales smaller than $2\pi k_\mathrm{s}^{-1}$ to evolve more slowly. However, since the amplitude is bound by the maximal value $\approx 54.78$, see (\ref{eq:19}), there is no truncation scale such that the free Zel'dovich power spectrum reaches the amplitude of the linear CDM power spectrum at small scales today.

We conclude that our results agree with the known fact that the truncated Zel'dovich power spectrum cannot be tuned to yield the amplitude of the linear CDM power spectrum or even the non-linear power spectrum at $z=0$ \cite[for example]{Schneider1995}. Additionally, we derived the amplitude that any truncated Zel'dovich power spectrum can maximally reach at small scales. 

We finally remark that the exact cancellation of the exponential damping term $\E^{-Q}$ in (\ref{eq:1}) visible in the analytic asymptotic expansion requires particular care in numerical evaluations of the free power spectrum $\mathcal{P}$. The small-scale behaviour of the exponent in the integrand of (\ref{eq:1}) needs to be accurately modelled at least to order $q^2$ to avoid artificial exponential growth or damping.

\section{Full asymptotic series and analytic toy model}
\label{sec:4}

We now proceed to derive an expression of the entire asymptotic series for $\mathcal{P}$ by applying a resummed version of Laplace's method to the one-dimensional integral of the radial coordinate $q$ in (\ref{eq:1}).

\subsection{Full asymptotic series}

As shown in\cite{Fulks1961}, asymptotic series expansions for $d$-dimensional Laplace integrals can be obtained without any smoothness conditions on the involved functions, if only their asymptotic expansions near the critical minimum exist. The complete expansion series is obtained by first transforming the integrand to spherical coordinates. Afterwards, the one-dimensional Laplace method is applied to the integral over the radial coordinate. Finally, the integrations over the angles occuring in each term in the asymptotic expansion has to be performed.

We adopt this procedure, but to obtain an asymptotic series for $\mathcal{P}$ as $k \to \infty$, we have to reorder the double series arising by Laplace's method. In our final step, we resum the series to arrive at a compact expression for all coefficients in the asymptotic series.

We return to (\ref{eq:1}), but this time write the momentum auto-correlation matrix as
\begin{equation}
  \hat C_{pp}\left(\vec q\,\right) = -\id{3}a_1(q)-\pi_\parallel a_2(q)\;,
\label{eq:23}
\end{equation} 
where $\pi_\parallel$ is the projector
\begin{equation}
  \pi_\parallel = \frac{\vec q\otimes\vec q}{q^2}
\label{eq:24}
\end{equation} 
parallel to $\vec q$ and the momentum auto-correlation functions
\begin{equation}
  a_1(q) = \frac{\xi_\psi'(q)}{q}\;,\quad
  a_2(q) = \xi_\psi''(q)-a_1(q)
\label{eq:25}
\end{equation}
are determined by derivatives of the auto-correlation function
\begin{equation}
  \xi_\psi(q) = \int_k\,P_\psi^\mathrm{(i)}(k)\,\E^{\I\vec k\cdot\vec q}
\label{eq:26}
\end{equation}
of the initial velocity potential $\psi$.

We align the wave vector $\vec k \neq 0$ in (\ref{eq:1}) with the $z$-axis and transform to spherical coordinates $(q,\varphi,\mu)$, where $\mu = q_z/q$ is the cosine of the angle enclosed by $\vec q$ and $\vec k$. Since the integration kernel is independent of the azimuthal angle $\varphi$, we can write
\begin{equation}
  \mathcal{P}(k,t) \sim
  2 \pi \E^{-Q} \int_{-1}^1 \D\mu
  \int_0^{q_\mathrm{max}}\D q \, q^2 \E^{- g_{qp}^2 k^2 \left[
    a_1(q) + \mu^2 a_2(q)
  \right]} \E^{\I k\mu q} \quad \mbox{as} \quad k \to \infty\;.
\label{eq:27}
\end{equation}
The $q$ integration can stop at the arbitrary upper bound $q_\mathrm{max}>0$ because the asymptotic behaviour of the integrand is determined by the exponent in the immediate neighbourhood of its minimum. As explained above, we can neglect the delta distribution resulting from integrating the Fourier phase over the unity subtracted from the integrand in (\ref{eq:1}).

The resulting asymptotic series is derived in Appendix \ref{sec:B}. The result is
\begin{equation}
  \mathcal{P}(k,t) \sim \sum_{m=0}^\infty
  \frac{\mathcal{P}^{(m)}(t)}{k^{3+2m}}
  \quad \mbox{as} \quad k \to \infty\;,
\label{eq:28}
\end{equation}
with the time-dependent coefficients
\begin{equation}
  \mathcal{P}^{(m)}(t) \coloneqq \frac{\pi^{3/2}}{(2m)!}
  \lim_{x\to0}\,\frac{\D^{2m}}{\D x^{2m}}\left\{
    \left(
      \frac{x^2/g_{qp}^2(t,0)}{a_1(x)+\frac{\sigma_1^2}{3}}
    \right)^{m+1} 
    \left.\frac{\D^m}{\D y^m}
      \frac{y^m\sqrt{R(x,y)}}{\E^{R(x,y)}}
    \right|_{y=y_0(x)}
  \right\}\;.
\label{eq:29}
\end{equation}
containing the definitions
\begin{equation}
  R(x,y) = \frac{x^2/g_{qp}^2(t,0)}{a_2(x)}\frac{y}{1+y}\quad\mbox{and}\quad
  y_0(x) = \frac{a_2(x)}{a_1(x)+\sigma_1^2/3}\;.
\label{eq:30}
\end{equation} 
The coefficients $\mathcal{P}^{(m)}(t)$ are therefore given by compact expressions. Using the asymptotic expansions of the functions $a_1(q)$ and $a_2(q)$ for $q\to0$ derived in Appendix \ref{sec:C}, the first two coefficients turn out to be
\begin{align}
  \mathcal{P}^{(0)}(t) &= 3\,(4\pi)^{3/2}\,\Sigma^{3/2}(t)\,\E^{-\Sigma(t)}\;,
  \nonumber\\
  \mathcal{P}^{(1)}(t) &= \frac{(4 \pi)^{3/2}}{28}\,
  \frac{\sigma_3^2}{\sigma_2^2}\,\Sigma^{5/2}(t)\,\E^{-\Sigma(t)}\left[
    123-132\Sigma(t)+20\Sigma^2(t)
  \right]\;.
 \label{eq:31}
\end{align}
In general, since the asymptotic series of $a_1(q)$ and $a_2(q)$ in the limit $q\to0$ are even in $q$ with the $n$-th coefficients being proportional to $\sigma_{n+1}^2$, (\ref{eq:31}) implies that $\mathcal{P}^{(m)}(t)$ depends only on moments of the initial velocity potential that are of order $m+2$ and lower. Furthermore,
\begin{equation}
  \mathcal{P}^{(m+1)}(t) \propto \sigma_{m+3}^2
  \quad\mbox{for all}\quad
  m = 0,1,2,\ldots\;,
\label{eq:32}
\end{equation}
showing that higher orders of the asymptotics series for the free power spectrum $\mathcal{P}$ carry higher orders of the initial velocity potential.

\subsection{Analytic toy model}

\begin{figure}
  \includegraphics[width=\hsize]{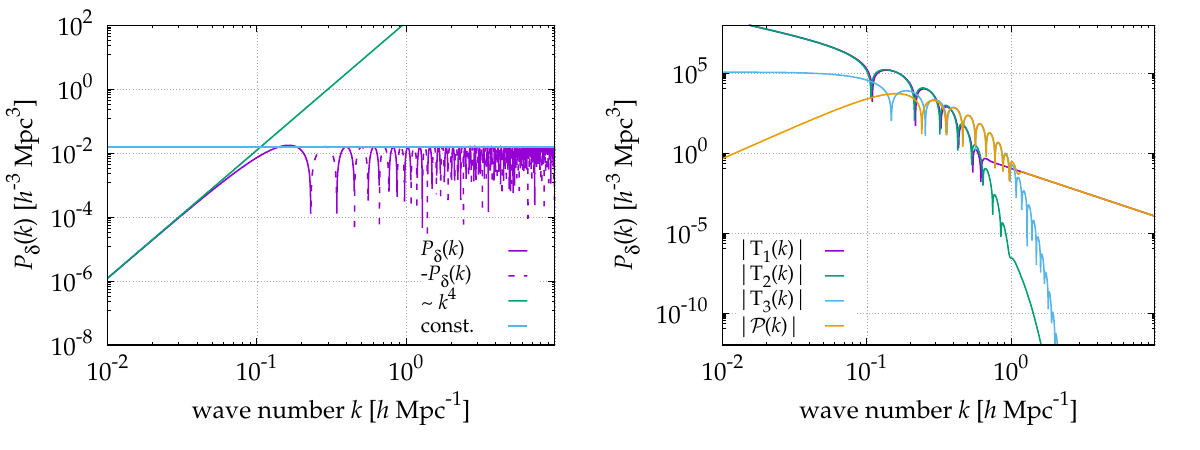}
\caption{Left: The toy-model power spectrum (\ref{eq:33}) increases $\propto k^4$ for small $k\ll k_0$. For large $k\gg k_0$, it oscillates like a sine wave, phase shifted by $-3\pi/2$. Dashed lines indicate negative values. Right: On large scales (small wave numbers), the three terms $T_{1,2,3}$ cancel to several orders. At intermediate scales, $\mathcal{P}$ approximately equals $T_3$, while $T_1$ dominates on small scales.}
\label{fig:5}
\end{figure}

The free power spectrum can be calculated analytically for the initial density power spectrum
\begin{equation}
  P_\delta^\mathrm{(i)}(k) = Akj_3(k)\;,
\label{eq:33}
\end{equation}
with amplitude $A$ and the dimension-less wave number $k$ referred to some arbitrary scale $k_0$ set to unity. This spectrum is shown in the left panel of Fig.~\ref{fig:5}. Clearly, this example is unphysical for cosmic structure formation because real fields, like the cosmic density fluctuations, cannot have negative-valued power spectra. Nevertheless, we find this example very instructive from a mathematical point of view, and worth discussing.

We show in App.~\ref{sec:D} that the spectrum (\ref{eq:33}) has the moment
\begin{equation}
  \frac{\sigma_1^2}{3} = \frac{A}{8\pi}\;,
\label{eq:34}
\end{equation}
and the correlation functions $a_1(q)$ and $a_2(q)$ are
\begin{equation}
  a_1(q) = \frac{\sigma_1^2}{3}\left(q^2-1\right)
  \Theta\left(1-q\right) \;,\quad
  a_2(q) = \frac{2\sigma_1^2}{3}\,q^2\,
  \Theta\left(1-q\right)
\label{eq:35}
\end{equation}
with the Heaviside step function $\Theta$. The radial coordinate $q$ is dimension-less as well, expressed in units of $k_0^{-1}$. With these expressions, aligning the wave vector once more with the $z$ axis, we compute the free power spectrum
\begin{align}
  \mathcal{P}(k,t) &= 2\pi\int_{-1}^1\D\mu\int_0^{k_0^{-1}}\D q\,q^2\left[
    \E^{-Q(1+2\mu^2)q^2}-\E^{-Q}
  \right]\E^{\I k\mu q} \nonumber\\ &=
  T_1(k)+T_2(k)+T_3(k)
\label{eq:36}
\end{align}
with $Q$ from (\ref{eq:3}). With the definition
\begin{equation}
  \Sigma  = \left(4\tau_1^2\right)^{-1}\;,
\label{eq:37}
\end{equation}
the three terms $T_{1,2,3}(k)$ resulting from the integration in (\ref{eq:53}) are
\begin{align}
  T_1(k) &= \frac{3(4\pi\Sigma)^{3/2}}{k^3}
  \E^{-\Sigma}\Re\erf\left(\tau_1k+\frac{\I}{2\tau_1}\right)\;, \nonumber\\
  T_2(k) &= -\frac{2(6\pi\Sigma)^{3/2}}{k^3}
    \E^{-3\Sigma/2-Q}\Re\erf\left[
      \frac{1}{\sqrt{6}}\left(
        2\tau_1k+\frac{3\I}{2\tau_1}
      \right)
    \right]\;,\quad\mbox{and} \nonumber\\
  T_3(k) &= -\frac{4\pi}{k}\,\E^{-Q}\,j_1(k)\;.
\label{eq:38}
\end{align}

The term $T_3$ falls off exponentially for $k\to\infty$. For $T_1$ and $T_2$, we recall $\erf z^* = \erf^*z$ for $z\in\mathbb{C}$ to write $2\Re\erf z = \erf z+\erf z^*$ and use the asymptotic expansion
\begin{equation}
  \erf(a k+\I b)+\erf(a k-\I b) = 2+\mathcal{O}\left(
    \frac{\E^{-a^2 k^2}}{k}\cos(2abk)
  \right)\;,
\label{eq:39}
\end{equation} 
from which we can conclude that $T_2$ is also exponentially suppressed. For $T_1$, we get
\begin{equation}
  T_1(k) \sim \frac{3(4\pi\Sigma)^{3/2}}{k^3}\,\E^{-\Sigma}
  \quad\mbox{as}\quad k\to\infty\;,
\label{eq:40}
\end{equation}
which is in perfect agreement with the asymptotics that we derived above.

Figure~\ref{fig:5} illustrates what happens to this power spectrum: Free streaming first suppresses power on small scales and leads to an exponential decrease of the power spectrum. The collective motion of the particles with correlated momenta wipes out the oscillatory pattern in the initial power spectrum and builds up the power on small scales.

\section{Summary and discussion}
\label{sec:5}

In Sect.~\ref{sec:2} of this paper, we derived the leading-order asymptotic behaviour on small scales, $k\to\infty$, of the free density-fluctuation power spectrum $\mathcal{P}$ in KFT, given by (\ref{eq:1}). The free power spectrum in KFT corresponds to the Zel'dovich power spectrum which is non-linear in the sense that arbiratrily large density fluctuations can occur in the Zel'dovich approximation. By applying rigorous results for $d$-dimensional integrals of the Fourier-Laplace type, derived in Appendix \ref{sec:A}, we proved that the free power spectrum always develops a $k^{-3}$ tail, irrespective of the cosmological model and the shape of the power spectrum characterizing the initial particle distribution. This small-scale behavior has been suggested before by \cite{Schneider1995} for the non-linear Zel'dovich power spectrum. Independently of our work, the $k^{-3}$ asymptotics has recently been derived by \cite{Chen2020} in a non-perturbative framework, finding the same amplitude for the Zel'dovich dynamics that we derive here.

After discussing cosmological consequences, we then derived the complete asymptotic series of $\mathcal{P}$. We proved that terms of order $m$ fall off asymptotically proprotional to $k^{-3-2m}$, derived a closed equation for the coefficients $\mathcal{P}^{(m)}$ of this series, and gave specific expressions for these coefficients in leading and next-to-leading order.

An important next step will be to study whether the $k^{-3}$ behavior at leading order, which we proved here for free trajectories, also persists when more gravitational interactions are included, for example via perturbation or mean-field theory. A Mean-field theory has been developed within kinetic field theory (see \cite{2021ScPP...10..153B}), where we could show that the non-linear power spectrum obtained with mean-field theory agrees with simulations within a few percent up to wave numbers of $k \approx 10\,h\,\mathrm{Mpc}$. Since the starting point of this non-linear power spectrum is the Zel'dovich power spectrum which is then multiplied by an averaged interaction term which has been shown to become scale independent for large wave numbers $k$, this indicates that the shape of the non-linear power spectrum on small scales has to be a property of the Zel'dovich power spectrum already, while particle interactions going beyond the Zel'dovich approximation do not change the shape of the power spectrum at large $k$. Additionally, we only took initial momentum-momentum correlations into account in this work, neglecting initial density-density and initial density-momentum correlations. At large scales, it has been shown that this is a valid approach. At small scales, however, this assumption also has to be studied more thoroughly. Furthermore, it will be subject to future work to analyze the origin of the observed universal dark matter halo density profiles, especially in their central regions.

It is an important implication of our work that the exponential damping factor in front of the integral (\ref{eq:1}) is exactly cancelled by the asymptotic expansion. Freely streaming, classical particles initially drawn from a Gaussian random density and velocity field will thus not lead to an exponential damping of small-scale structures, but to structures with a dimension-less power spectrum $k^3\mathcal{P}$ developing a constant level for $k\to\infty$. Apart from its physical significance, this cancellation of the exponential damping term also has implications for the implementation of codes for evaluating the free power spectrum. It is crucial for the evaluation at large wave numbers that the small-scale behavior of the initial momentum-correlation function is implemented asymptotically correctly up to order $q^2$. The results will otherwise diverge artificially and exponentially.

We further showed that the time evolution of the amplitude of the leading-order, $k^{-3}$ asymptotic tail depends only on the product $g_{qp}^2(t,0)\sigma_2^2$, showing that the evolution of the small-scale structure proceeds more slowly when the initial power spectrum is normalized to have a smaller moment $\sigma_2^2$. At early times, the amplitude rises steeply, indicating the convergence of particle streams due to their initially correlated momenta. The leading-order asymptotic term then reaches a universal maximum value $\approx 54.78$ when $g_{qp}^2(t,0)\sigma_2^2 = 5/3$. This maximum amplitude marks small-scale stream crossing, where particle trajectories cross and the particles subsequently move apart. This leads to the re-expansion of the small-scale structures that is known to occur in the Zel'dovich approximation and that goes along with the decreasing amplitude at small scales.

With this, we lay the foundation for future work to explore the possible imprint of these scales in observables and to derive the precise relations for specific dark matter candidates.

\section*{Acknowledgements}
We thank Manfred Salmhofer for numerous helpful discussions and for reviewing a draft of this paper. Constructive comments by Robert Lilow, Ricardo Waibel, Leif Seute and Hannes Riechert contributed to clarifying the line of reasoning and improving numerical algorithms. This work is funded by the Deutsche Forschungsgemeinschaft (DFG, German Research Foundation) under Germany's Excellence Strategy EXC 2181/1 - 390900948 (the Heidelberg STRUCTURES Excellence Cluster). 

\section*{Data Availability}
The data underlying this article will be shared on reasonable request to the corresponding author.

\bibliographystyle{mnras}
\bibliography{main}

\appendix
\section{Asymptotics of $d$-dimensional Laplace-Fourier integrals}
\label{sec:A}

We consider $d$-dimensional Laplace-Fourier integrals
\begin{equation}
  P(k) = \int_\Omega\E^{-|k|^sf(x)}\E^{\I k \cdot x}\D^d x
\label{eq:41}
\end{equation}
where the vector $k \in \mathbb{R}^d$ appears with different powers in the real and the imaginary exponent of the integrand. For this type of integrals, we wish to study the asymptotics in the limit $|k| \to \infty$. To this end, we need to prepare some concepts and notation. Here and below, $D$ denotes the derivative of a function with respect to its arguments.

\subsection{Preliminary remarks}

Let $\Omega\subset\mathbb{R}^d$ be an open subset of $\mathbb{R}^d$ and $x_0\in\Omega$ a non-degenerate critical point of the smooth function $f:\Omega\to\mathbb{R}$ on $\Omega$ with $f\in C^\infty$. Then, by Morse's lemma, neighbourhoods $U$, $V$ of the points $y = 0$ and $x = x_0$ and a diffeomorphism $h: U\to V$ exist such that
\begin{equation}
  \left(f\circ h\right)(y) = f(x_0)+\frac{1}{2}y^\top Qy\;,\quad
  Q = \mathrm{diag}\left(\mu_1,\ldots,\mu_d\right)\;,
\label{eq:42}
\end{equation}
where the Jacobian $H := Dh$ of $h$ has unit determinant at $y = 0$, $\det H_0= 1$ with $H_0 := H(0)$. Then, by the chain rule and $(Df)(x_0) = 0$, $H_0$ diagonalises the Hessian matrix
\begin{equation}
  A := \left(D^2f\right)\Bigr\vert_{x=x_0}
\label{eq:43}
\end{equation}
of $f$ in $x_0$ such that
\begin{equation}
  Q = H^\top_0AH_0\;.
\label{eq:44}
\end{equation}

With the same chart $h:U\to V$, the Jacobian $H=Dh$ and a smooth function $g:\Omega\to\mathbb{R}$ with $g\in C^\infty$, we define the function $G:U\to\mathbb{R}$ by $G(y) = (g\circ h)(y)\det H(y)$. We further introduce multi-indices $\alpha=(\alpha_1, \alpha_2, \ldots, \alpha_d)$ and agree on the notation
\begin{align}
  |\alpha| &:= \alpha_1+\alpha_2+\ldots+\alpha_d\;,\nonumber\\
  \alpha!  &:= \alpha_1!\alpha_2!\cdots\alpha_d!\;,\nonumber\\
  \Gamma(\alpha) &:=
  \Gamma(\alpha_1)\Gamma(\alpha_2)\cdots\Gamma(\alpha_d)\;,\nonumber\\
  \mu^\alpha &:= \mu_1^{\alpha_1}\mu_2^{\alpha_2}\cdots\mu_d^{\alpha_d}\;,
\label{eq:45}
\end{align}
where $\mu = (\mu_1, \mu_2, \ldots, \mu_d)$ is a $d$ dimensional real-valued vector.
We finally define the symbol
\begin{equation}
  \delta(\alpha) = \begin{cases}
                     1 & \mbox{all $\alpha_j$ even} \\
                     0 & \mbox{else}
                   \end{cases}
\label{eq:46}
\end{equation}
and the derivative operator
\begin{equation}
  D^\alpha G(0) = \frac
  {\partial^{|\alpha|}}{\partial^{\alpha_1}y_1\ldots\partial^{\alpha_d}y_d}
  G(y)\Biggr\vert_{y = 0}\;.
\label{eq:47}
\end{equation}
With these preparations, we can now continue with the following theorem for the asymptotic behaviour of a class of integrals on $\Omega$ \cite{UBHD-65467105}.

\begin{theorem}
\label{the:2}
  Let $J(\lambda)$ be the integral
  \begin{equation}
    J(\lambda) = \int_\Omega\E^{-\lambda f(x)}g(x)\,\D x
  \label{eq:48}
  \end{equation}
  with $x\in\mathbb{R}^d$ and $\Omega\subset\mathbb{R}^d$, whose asymptotic behaviour we wish to determine for $\lambda\to\infty$. We assume that
  \begin{enumerate}
    \item  $f, g\in C^\infty$ on $\Omega$;
    \item $J(\lambda)$ converges absolutely for sufficiently large $\lambda$;
    \item $f$ has a minimum at, and only at, $x_0\in\Omega$ such that
    \begin{equation}
      \rho(\varepsilon) := \inf_{\Omega\setminus B_\varepsilon(x_0)}f(x)-f(x_0) > 0
    \label{eq:49}
    \end{equation}
    for all $\varepsilon>0$, where $B_\varepsilon(x_0)$ is the open ball with radius $\varepsilon$ around $x_0$; and that
    \item the Hessian matrix $A$ of $f$ in $x_0$ is positive definite.
  \end{enumerate}
  Then, the integral $J(\lambda)$ has the asymptotic expansion
  \begin{equation}
    J(\lambda) \sim \E^{-\lambda f(x_0)}\sum_{n=0}^\infty
    \frac{c_n}{\lambda^{d/2+n}}
  \label{eq:50}
  \end{equation}
  for $\lambda\to\infty$, with
  \begin{equation}
    c_n = \sum_{|\alpha|=2n}\delta(\alpha)
    \left(\frac{2}{\mu}\right)^{(\alpha+1)/2}
    \Gamma\left(\frac{\alpha+1}{2}\right)\frac{D^\alpha G(0)}{\alpha!}\;.
  \label{eq:51}
  \end{equation}
  The $\mu_i$, $1\le i\le d$, are the (real and positive) eigenvalues of $A$.
\end{theorem}

\subsection{Laplace-Fourier integrals in $d$ dimensions} \label{sec:a2}

We now apply the result (\ref{eq:50}) with the coefficients (\ref{eq:51}) to Laplace-Fourier-type integrals of the form
\begin{equation}
  J(\lambda, k) = \int_\Omega\E^{-\lambda f(x)}\E^{\I kx}\D^d x\;,
\label{eq:52}
\end{equation}
whose asymptotic behaviour we wish to determine for $\lambda\to\infty$, where $|k|$ is assumed to be a large parameter at the same time. Without loss of generality, we set $x_0 = 0$ here. For integrals of this type, the function $G$ defined above is
\begin{equation}
  G(y) = \E^{\I kh(y)}\det H(y)\;.
\label{eq:53}
\end{equation}
Since only the derivatives of $G$ acting on the exponential factor increase the powers of $k$, and since additionally $G(0) = 1$, we can state
\begin{equation}
  D^\alpha G(0) \sim
  \prod_{j=1}^d\left[\I\left(H^\top_0k\right)_j\right]^{\alpha_j} =
  \I^{|\alpha|}
  \prod_{j=1}^d\left(H^\top_0k\right)_j^{\alpha_j}
  \quad\mbox{for}\quad |k|\to\infty
\label{eq:54}
\end{equation}
and move on to write
\begin{equation}
  c_n(k) \sim \sum_{|\alpha|=2n}\delta\left(\alpha\right)\left(-1\right)^n
  \prod_{j=1}^d\left(\frac{2}{\mu_j}\right)^{(\alpha_j+1)/2}
  \Gamma\left(\frac{\alpha_j+1}{2}\right)
  \frac{\left(H^\top_0k\right)_j^{\alpha_j}}{\alpha_j!}
  \quad\mbox{for}\quad |k|\to\infty\;.
\label{eq:55}
\end{equation}
Since the factor $\delta(\alpha)$ ensures that only even $\alpha_j$ can occur in the sum, it is convenient to define a multi-index $n$ with elements $n_j=\alpha_j/2$. Evaluating the Gamma functions for odd arguments in (\ref{eq:55}) and noting that $\prod\mu_j = \det A$, we arrive at
\begin{equation}
  c_n(k) \sim \sqrt{\frac{(2\pi)^d}{\det A}}\left(-\frac{1}{2}\right)^n
  \sum_{|n|=n}\prod_{j=1}^d\frac{1}{n_j!}
  \frac{\left(H^\top_0k\right)_j^{2j}}{\mu_j^{n_j}}
  \quad\mbox{for}\quad |k|\to\infty\;.
\label{eq:56}
\end{equation}

Since (\ref{eq:44}) implies that $H_0Q^{-1}H_0^\top = A^{-1}$, the remaining sum equals
\begin{equation}
  \sum_{|n|=n}\prod_{j=1}^d\frac{1}{n_j!}
    \frac{\left(H^\top_0k\right)_j^{2j}}{\mu_j^{n_j}} =
  \frac{1}{n!}
  \left[
    \left(H^\top_0k\right)^\top Q^{-1}\left(H^\top_0k\right)
  \right]^n = \frac{\left(k^\top A^{-1}k\right)^n}{n!}
\label{eq:57}
\end{equation}
allowing us to write (\ref{eq:50}) in the form
\begin{equation}
  J(\lambda,k) \sim \E^{-\lambda f(0)}\sqrt{\frac{(2\pi)^d}{\lambda^d\det A}}
  \exp\left(-\frac{k^\top A^{-1}k}{2\lambda}\right)
  \quad\mbox{for}\quad \lambda\to\infty\;.
\label{eq:58}
\end{equation}

For integrals of the type
\begin{equation}
  P(k) = \int_\Omega\E^{-|k|^sf(x)}\E^{\I kx}\D^d x
\label{eq:59}
\end{equation}
with $s\ge2$, we substitute again $\lambda\to|k|^s$ in (\ref{eq:53}). We then obtain immediately from (\ref{eq:58})
\begin{equation}
  P(k) \sim \E^{-|k|^sf(0)}
  \sqrt{\frac{(2\pi)^d}{|k|^{sd}\det A}}
  \exp\left(-\frac{k^\top A^{-1}k}{2|k|^s}\right)
  \quad\mbox{for}\quad |k|\to\infty\;,
\label{eq:60}
\end{equation}
which proves the result (\ref{eq:3}).

\subsection{Extension to $\mathbb{R}^d$} \label{sec:a3}

We finally wish to extend the integration domain $\Omega$ to $\mathbb{R}^d$. We begin with the conditions imposed in Theorem \ref{the:2}, but relax conditions (1) and (2) to
\begin{enumerate}
  \item[$1'$.] the Hessian $A=D^2f$ exists in $x_0 = 0$ and
  \item[$2'$.] $f\in L_2(\mathbb{R}^d)$, i.e. $f$ is quadratically integrable in $\mathbb{R}^d$
\end{enumerate}
and add as a further condition:
\begin{enumerate}
  \item[3b.] The infimum
  \begin{equation}
    \sigma(\varepsilon) := \inf_{\mathbb{R}^d\setminus B_\varepsilon(0)}|f(0)|-|f(x)| > 0
  \label{eq:61}
  \end{equation}
  for all $\varepsilon>0$.
\end{enumerate}
Note that conditions (3) and (3b) together imply that $f(0)<0$. Then, the integral
\begin{equation}
  P(k) = \int_{\mathbb{R}^d}\E^{-|k|^sf(x)}\E^{\I kx}\D^d x
\label{eq:62}
\end{equation}
has the asymptotic expansion (\ref{eq:60}).

To prove this statement, we consider the integral
\begin{align}
  \bar P(k) :&= \int_{\mathbb{R}^d}\left[\E^{-|k|^sf(x)}-1+|k|^sf(x)\right]
  \E^{\I kx}\D^d x \nonumber\\ &=
  P(k)-(2\pi)^N\delta_\mathrm{D}(k)+|k|^s\tilde f(k) \nonumber\\ &=
  P(k)+|k|^s\tilde f(k)
\label{eq:63}
\end{align}
for $k>0$, where $\tilde f$ denotes the Fourier transform of $f$. We further split the integration domain as
\begin{equation}
  \bar P(k) = \left(\int_\Omega+\int_{\mathbb{R}^d\setminus\Omega}\right)
  \left[
    \E^{-|k|^sf(x)}-1+|k|^sf(x)
  \right]\E^{\I kx}\D^d x =: \bar P_\Omega(k)+\bar P_{\bar\Omega}(k)\;.
\label{eq:64}
\end{equation}
For the asymptotic expansion of $\bar P_\Omega(k)$, we estimate
\begin{equation}
  \bar P_\Omega(k) \le \int_\Omega\E^{-|k|^sf(x)}\E^{\I kx}\D^d x+\left[
    1+|k|^s|f(0)|
  \right]V_\Omega\;,
\label{eq:65}
\end{equation} 
where $V_\Omega$ is the volume of the domain $\Omega$. Since the second term on the right-hand side of (\ref{eq:65}) is exponentially suppressed compared to the asymptotic expansion (\ref{eq:60}) of the first term (recall that $f(0)<0$), we can conclude that
\begin{equation}
  \bar P_\Omega(k) \sim \E^{-|k|^sf(0)}\sqrt{\frac{(2\pi)^d}{|k|^{sd}\det A}}
  \exp\left(-\frac{k^\top A^{-1}k}{2|k|^s}\right)
  \quad\mbox{for}\quad |k|\to\infty\;.
\label{eq:66}
\end{equation} 
For $\bar P_{\bar\Omega}(k)$, we estimate
\begin{align}
  \bar P_{\bar\Omega}(k) &\le \sum_{n=2}^\infty\frac{|k|^{ns}}{n!}
  \int_{\mathbb{R}^d\setminus\Omega}\left|f(x)\right|^n\D^d x \nonumber\\ &\le
  \sum_{n=2}^\infty\frac{|k|^{ns}}{n!}\left[
    |f(0)|-\sigma(\varepsilon)
  \right]^{n-2}\underbrace{\int_{\mathbb{R}^d}|f(x)|^2\D^d x}_{=:\,C\,<\,\infty}
  \nonumber\\ &=
  \frac{C}{\left[
    |f(0)|-\sigma(\varepsilon)
  \right]^2}\left\{
    \E^{-|k|^s\left[f(0)+\sigma(\varepsilon)\right]}-1+|k|^s\left[
      f(0)+\sigma(\varepsilon)
    \right]
  \right\} \nonumber\\ &\sim
  \frac{C}{\left[
    |f(0)|-\sigma(\varepsilon)
  \right]^2}\E^{-|k|^s\left[f(0)+\sigma(\varepsilon)\right]}
  \quad\mbox{for}\quad |k|\to\infty\;.
\label{eq:67}
\end{align}
Again, this result is exponentially suppressed by the factor $\exp(-|k|^s\sigma(\varepsilon))$ compared to the asymptotic expansion (\ref{eq:60}). Since $\tilde f\in L_2(\mathbb{R}^d)$, we can conclude
\begin{align}
  P(k) &= \bar P(k)-|k|^s\tilde f(k) \sim \bar P(k) \sim \bar P_\Omega(k)
  \nonumber\\ &\sim
  \E^{-|k|^sf(0)}\sqrt{\frac{(2\pi)^d}{|k|^{sd}\det A}}
  \exp\left(-\frac{k^\top A^{-1}k}{2|k|^s}\right)
  \quad\mbox{for}\quad |k|\to\infty\;,
\label{eq:68}
\end{align}
as claimed.

\section{Asymptotics of a specific Laplace-Fourier type integral}
\label{sec:B}

We now wish to study the asymptotic behavior of the specific Laplace-Fourier type integral
\begin{equation}
  P(k) = \int_{-1}^{1} \mathrm{d}\mu \int_0^{\infty} \D q \; q^{2} \E^{-k^2 g_{qp}^2 \left[a_1(q)+ \mu^2 a_2(q)\right]}\E^{\I k \mu q}
\label{eq:69}
\end{equation}
for $k\to\infty$. To this end, we begin with an asymptotic expansion of the $q$ integral, for which we modify Erdélyi's theorem \cite{UBHD-65249191, UBHD-65249153, 1961ArRMA...7....1E} for Laplace integrals, which states:
\begin{theorem}[Erdélyi]
\label{the:1}
  Let $I(\lambda)$ be an integral of the form
  \begin{equation}
    I(\lambda) = \int_a^b\E^{-\lambda f(x)}g(x)\,\D x\;,
  \label{eq:70}
  \end{equation}
  where $f(x)$ is a real function of the real variable $x$, while $g(x)$ may be real or complex. Then, if
  \begin{enumerate}
    \item $f(x)>f(a)$ for $x\in(a,b)$ and
    \begin{equation}
      \inf_{[a+\delta,b)}f(x)-f(a)>0
    \label{eq:71}
    \end{equation}
    for $\delta>0$;
    \item $f'(x)$ and $g(x)$ are continuous in a neighbourhood of $a$, except possibly at $a$;
    \item $f$ and $g$ admit asymptotic expansions
    \begin{align}
      f(x) &\sim f(a)+\sum_{k=0}^\infty a_k(x-a)^{k+\alpha}\;,\nonumber\\
      g(x) &\sim \sum_{k=0}^\infty b_k(x-a)^{k+\beta-1}\;,
    \label{eq:72}
    \end{align}
    $f$ can be term-wise differentiated,
    \begin{equation}
      f'(x) \sim \sum_{k=0}^\infty a_k(k+\alpha)(x-a)^{k+\alpha-1}
    \label{eq:73}
    \end{equation}
    for $x\to a^+$, where $\alpha>0$ and $\mathrm{Re}\,\beta>0$; and
    \item $I(\lambda)$ converges absolutely for sufficiently large $\lambda$; then
  \end{enumerate}
  the integral $I(\lambda)$ has the asymptotic expansion
  \begin{equation}
    I(\lambda) \sim \E^{-\lambda f(a)}\sum_{n=0}^\infty
    \frac{\Gamma(\nu)\,c_n}{\lambda^\nu}
  \label{eq:74}
  \end{equation}
  for $\lambda\to\infty$, where $\nu := (n+\beta)/\alpha$. The coefficients $c_n$ can be expressed by $a_n$ and $b_n$ as
  \begin{equation}
    c_n = \frac{1}{\alpha a_0^\nu}\sum_{m=0}^n
    \frac{b_{n-m}}{m!}\,d_{m,n}\quad\mbox{with}\quad
    d_{m,n} = \lim_{x\to0}\frac{\D^m}{\D x^m}\left(
      1+\sum_{j=1}^\infty\frac{a_j}{a_0}\,x^j
    \right)^{-\nu}\;.
  \label{eq:75}
  \end{equation}
\end{theorem}

For applying Erdelyi's theorem, we set $\lambda := k^2 g_{qp}^2$ and define
\begin{equation}
  f_\mu(q) := a_1(q)+\mu^2 a_2(q)
  \quad \text{and} \quad
  g_{\mu,k}(q) := q^2 \E^{\I k\mu q}\;.
\label{eq:76}
\end{equation}
As we shall see in the next section, this function has the asymptotic expansion
\begin{equation}
  f_\mu(q) \sim -\frac{\sigma_1^2}{3}+\sum_{m=0}^\infty a_{2m}(\mu)\,q^{2m+2}
\label{eq:77}
\end{equation}
with
\begin{equation}
  a_{2m}(\mu) = \frac{(-1)^{m+2}\sigma_{m+2}^2}{(5+2m)(3+2m)!}\left(
    1+2(m+1)\mu^2
    \right)\;,
\label{eq:78}
\end{equation}
and the asymptotic expansion of the function $g_{\mu,k}(q)$ is
\begin{equation}
  g_{\mu,k}(q) \sim \sum_{m=0}^{\infty} b_m(\mu,k)\,q^{m+3-1}
  \quad \mbox{with} \quad
  b_m(\mu,k) = \frac{(\I k \mu)^m}{m!}\;.
\label{eq:79}
\end{equation}
We identify $\alpha = 2$ and $\beta = 3$ and find from (\ref{eq:70}) and (\ref{eq:74})
\begin{align}
  I_{\mu, k} (\lambda) &:= \int_0^{\infty}\D q\,q^{2}
  \E^{-\lambda f_\mu(q)}\E^{\I k \mu q} \nonumber\\ &\sim
  \int_0^{b} \D q \; \E^{-\lambda f_\mu(q)} g_{\mu,k}(q) \sim
  \E^{-\lambda f_\mu(0)}\sum_{n=0}^{\infty}
  \frac{\Gamma\left(\frac{n+3}{2}\right)c_n(\mu,k)}{\lambda^{(n+3)/2}}\;.
\label{eq:80}
\end{align}
The coefficients $c_n$ are
\begin{equation}
  c_n(\mu,k) = \frac{a_0^{-(n+3)/2}(\mu)}{2}
  \sum_{m=0}^n\frac{(\I k\mu)^{n-m}}{m!(n-m)!}d_{m,n}(\mu)
\label{eq:81}
\end{equation}
with
\begin{align}
  d_{m,n}(\mu) &= \lim_{x\to0}\frac{\D^m}{\D x^m}\left(
    1+\sum_{j=0}\frac{a_{2j}(\mu)}{a_0(\mu)}x^{2j}
  \right)^{-\frac{n+3}{2}} \nonumber\\ &= a_0^{(n+3)/2}(\mu)
  \lim_{x\to0}\frac{\D^m}{\D x^m}\left[
    \frac{f_\mu(x)-f_\mu(0)}{x^2}
  \right]^{-\frac{n+3}{2}}\;.
\label{eq:82}
\end{align}

Since $f_\mu(q)$ is even in $q$ (see also Sect.~\ref{sec:C} below), all odd derivatives vanish, implying $d_{2m+1,n}(\mu) = 0$. Thus in the coefficients $c_n(\mu,k)$, only even values of $m$ contribute to the sum. We further add that, due to the integration over the parameter $\mu$, it is sufficient to consider only the real part, which is even in $\mu$, i.e.\ only terms for even numbers of $n$. This brings us to
\begin{align}
  \Re I_{\mu, k}(\lambda) &\sim
  \frac{\E^{\lambda\frac{\sigma_1^2}{3}}}{2}\,
  \sum_{n=0}^{\infty}\frac{\Gamma\left(n+\frac{3}{2}\right)}{\lambda^{n+3/2}}\,
  \sum_{m=0}^n\frac{(\I k\mu)^{2n-2m}}{(2m)!(2n-2m)!} 
  \lim_{x\to0}\frac{\D^{2m}}{\D x^{2m}}\left[
    \frac{f_\mu(x)-f_\mu(0)}{x^2}
  \right]^{-n-\frac{3}{2}}\;.
\label{eq:83}
\end{align}

This is an asymptotic expansion of the integral for $\lambda \to \infty$. Note that this is not an asymptotic series for $k \to \infty$, which can be seen as follows. Each term in the sum over $n$ is a finite sum over $m$. Considering for each $n$ the $m=0$ term in the second sum, we observe that terms of order $k^3$ are contained in each $n$ term. If, however, the asymptotic series above was an asymptotic series for $k \to \infty$, then the terms of the first sum would have to contain terms of descending order in $k$. To get the first term of the asymptotic series in $k$, we therefore have to to collect all terms in the total expression that are of order $k^{-3}$. For the second term, we have to collect all $k^{-5}$-terms and so on. The result of this ordering, together with substituting $\lambda\to g^2_{qp} k^2$, is the asymptotic series
\begin{equation}
  \Re I_{\mu, k}\left(g_{qp}^2k^2\right) \sim
  \frac{\E^{g_{qp}^2k^2\frac{\sigma_1^2}{3}}}{2k^3}
  \sum_{m=0}^{\infty}\frac{(-k^2\mu^2)^{-m}}{(2m)!}\tilde{c}_m(\mu^2)
\label{eq:84}
\end{equation}
for $k\to\infty$ with the coefficients
\begin{equation}
  \tilde{c}_m(\mu^2) := \lim_{x \to 0}\frac{\D^{2m}}{\D x^{2m}}\left[
    \frac{x^2/g_{qp}^2}{f_\mu(x)-f_\mu(0)}
  \right]^{\frac{3}{2}}\sum_{n=0}^{\infty}
  \frac{\Gamma\left(n+m+\frac{3}{2}\right)}{(2n)!}\left[
    \frac{-\mu^2x^2/g_{qp}^2}{f_\mu(x)-f_\mu(0)}
  \right]^{n+m}\;.
\label{eq:85}
\end{equation}

We now proceed by performing the $\mu$-integral for each individual term,
\begin{equation}
  \int_{-1}^1\D\mu\,\left(-\mu^2\right)^{-m}\tilde{c}_m(\mu^2) =
  \sqrt{\pi}\lim_{x \to 0}\frac{\D^{2m}}{\D x^{2m}}\left(
    \frac{x^2/g_{qp}^2}{a_1(x)+\frac{\sigma_1^2}{3}}
  \right)^{m+1}\frac{\D^m}{\D y^m} y^{m}\frac{\sqrt{R(x,y)}}{\E^{R(x,y)}}\;,
\label{eq:86}
\end{equation}
to find the asymptotic expansion for the power spectrum
\begin{align}
  P(k) &= \int_{-1}^{1}\D\mu\int_0^{\infty}\D q\,q^2
  \E^{-k^2 g_{qp}^2f_\mu(q)}\E^{\I k\mu q} \nonumber\\ &\sim 
  \sqrt{\pi}\frac{\E^{k^2\frac{\tau_1^2}{3}}}{2k^3}
  \sum_{m=0}^\infty \frac{k^{-2m}}{(2m)!}
  \lim_{x \to 0}\frac{\D^{2m}}{\D x^{2m}}\left(
    \frac{x^2/g_{qp}^2}{a_1(x)+\frac{\sigma_1^2}{3}}
  \right)^{m+1}\frac{\D^m}{\D y^m}y^{m} 
  \frac{\sqrt{R(x,y)}}{\E^{R(x,y)}}\;.
\label{eq:87}
\end{align}
The function $R(x,y)$ is defined in (\ref{eq:30}).

\section{Asymptotic expansion of the functions $a_{1,2}(q)$ and $f_\mu(q)$}
\label{sec:C}

The functions $a_{1,2}(q)$ defined in (\ref{eq:4}) in terms of the correlation function $\xi_\psi(q)$ (\ref{eq:5}) of the velocity potential can be written as
\begin{equation}
  a_1(q) = -\frac{1}{2\pi^2}\int_0^\infty\D k\,P_\delta^\mathrm{(i)}(k)\,
  \frac{j_1(kq)}{kq}\;,\quad
  a_2(a) = qa_1'(q)\;.
\label{eq:88}
\end{equation}
Inserting the Taylor series for the first-order spherical Bessel function $j_1(kq)$ leads to the asymptotic series
\begin{align}
  a_1(q) &\sim \sum_{n=0}^\infty
  \frac{(-1)^{n+1}\sigma_{n+1}^2}{(3+2n)(1+2n)!}\,q^{2n}\quad\mbox{and}
  \nonumber\\
  a_2(q) &\sim \sum_{n=1}^\infty
  \frac{(-1)^{n+1}\,2n\,\sigma_{n+1}^2}{(3+2n)(1+2n)!}\,q^{2n}
  \quad\mbox{as}\quad q\to0\;.
\label{eq:89}
\end{align}
The function $f_\mu(q)$ from (\ref{eq:76}) thus has the asymptotic series
\begin{equation}
  f_\mu(q) = -\frac{\sigma_1^2}{3}+\sum_{m=0}^\infty
  \frac{(-1)^{m+2}\sigma_{m+2}^2}{(5+2m)(3+2m)!}
  \left(1+2(m+1)\mu^2\right)\,q^{2m+2}\;.
\label{eq:90}
\end{equation}
Comparing to (\ref{eq:72}), we can read off $\alpha = 2$ and
\begin{equation}
  a_0(\mu) = \frac{\sigma_2^2}{30}\left(1+2\mu^2\right)\;,\quad
  a_1(\mu) = 0\;,\quad
  a_2(\mu) = -\frac{\sigma_3^2}{840}\left(1+4\mu^2\right)\;.
\label{eq:91}
\end{equation} 

\section{Analytic toy model -- calculations}
\label{sec:D}

We use the identity
\begin{equation}
  \int_0^\infty\D x\,x^{\nu-\mu+1/2}\sqrt{xy}\,J_\mu(ax)\,J_\nu(xy) =
  \frac{2^{\nu-\mu+1}y^{\nu+1/2}}{\Gamma(\mu-\nu)a^\mu}
  \left(a^2-y^2\right)^{\mu-\nu-1}\,\Theta(a-y)
\label{eq:92}
\end{equation} 
valid for $-1 < \Re\nu < \Re\mu$ and $a > 0$ to calculate the correlation functions $a_{1,2}(q)$ for the power spectrum (\ref{eq:33}). Setting $\nu=3/2$ or $\nu=5/2$, $\mu = 7/2$, $a = k_0^{-1}$ and $y=q$ in (\ref{eq:92}) results in
\begin{equation}
  a_1(q) = -\frac{1}{2\pi^2}\int_0^\infty\D k\,P_\delta^{(i)}(k) 
  \frac{j_1(kq)}{kq} =
  \frac{A}{8\pi}\,\left(q^2-1\right)\,\Theta(1-q)
\label{eq:93}
\end{equation}
and
\begin{equation}
  a_2(q) = \frac{1}{2\pi^2}\int_0^\infty\D k\,P_\delta^{(i)}(k)
  j_2(kq) = \frac{A}{4\pi}\,q^2\,\Theta(1-q)\;.
\label{eq:94}
\end{equation}
Using
\begin{equation}
  \lim_{q\to0}\,\frac{j_1(kq)}{kq} = \frac{1}{3}
\label{eq:94a}
\end{equation}
in (\ref{eq:93}), we immediately conclude
\begin{equation}
  \frac{\sigma_1^2}{3} = -a_1(q=0) = \frac{A}{8\pi}\quad\Rightarrow\quad
  A = \frac{8\pi\sigma_1^2}{3}\;.
\label{eq:95}
\end{equation}

\bsp	
\label{lastpage}
\end{document}